\documentclass[conference,letter,letterpaper]{IEEEtran}

\makeatletter
\newtheorem{thm}{Theorem}

\usepackage{url}

\usepackage{ifpdf}
 \ifpdf
\usepackage[pdftex]{graphicx}
 \else
\usepackage[dvips]{graphicx}
 \fi

\ifCLASSINFOpdf
\else
\fi
\begin{document}
%
\def\mytitle{
On Subversive Miner Strategies 
and Block Withholding Attack 
in Bitcoin Digital Currency 
}
\title{\mytitle\\
\vskip-8pt
}

\author{\IEEEauthorblockN{
Nicolas T. Courtois
}
\IEEEauthorblockA{
University College London, UK
}
\and
\IEEEauthorblockN{Lear Bahack}
\IEEEauthorblockA{Open University of Israel}
}

%


\maketitle

\begin{abstract}
Bitcoin is a ``crypto currency'',
a decentralized 
electronic 
payment scheme based on cryptography.
Bitcoin economy grows at an incredibly fast rate
and is now worth some 10 billions of dollars.
Bitcoin mining is an activity which consists of creating (minting)
the new coins which are later put into circulation.
Miners spend electricity on solving cryptographic puzzles and
they are also gatekeepers which validate bitcoin transactions of other people.
Miners are expected to be honest and have some incentives to behave well.
However. In this paper we look at the miner strategies
with particular attention paid to subversive and dishonest strategies
or those which could put bitcoin and its reputation in danger.
We study in details several recent attacks
in which dishonest miners obtain
a higher reward than their relative contribution to the network.
In particular we revisit the concept of block withholding attacks
and propose a new concrete and practical block withholding attack
which we show to maximize the advantage gained by rogue miners.

{\bf RECENT EVENTS:}
It seems that our attack was executed in practice
against Eligius mining pool,
see Section \ref{RecentWithholdAttackEligius300BTC}.
\end{abstract}


 \begin{center}
 {\bf Keywords:}
electronic payment,
crypto currencies,
bitcoin,
bitcoin mining,
mining pools,
game theory
 \end{center}
%
\IEEEpeerreviewmaketitle

\section{Introduction}

Bitcoin digital currency \cite{SatoshiPaper} is not properly speaking a currency.
It is first of all an electronic payment system based on cryptography.
The fact that in cryptography we call it a ``crypto currency''
or ``e-money'' does not mean that it has to have
all the attributes of money,
or it should work as a replacement of traditional fiat currencies.
On the contrary. Bitcoin is rather
an emerging technology ecosystem and a social experiment,
which is still largely under development.
A paper at the Financial Cryptography 2012 conference explains that
Bitcoin is a system which
{\em uses no fancy cryptography, and is by no means perfect}
\cite{BitcoinFC12SecurityOverview}.
Bitcoin is a sort of financial anarchy.
A self-governing open-source crypto co-operative
which initially concerned only a few enthusiasts.
However in 2013 the press and the media
have popularized bitcoin and given it some serious attention.
In April 2013 the Economist have
explained that bitcoin is certainly
is one of the things which are going
to shape the future of finance and payment
\cite{TheEconomistDigitalGold}
and famously compared bitcoin to \emph{digital gold}.
Since this moment the market price of bitcoin
has increased nearly 8 times
with a particularly rapid increase at the end of 2013. 

\subsection{Bitcoin As A Distributed System}

Bitcoin is a decentralized peer-to-peer system
with a network of open-source software nodes
\cite{BitcoinMainSoftwareDistribution}.
The network is run by the interested
participants: 
people who use it to make payments between themselves.
As such it requires no trusted parties such as
traditional financial institutions.
Bitcoin is build with the idea that -- maybe--
we do not need trust and good reputation.
Neither we would need regulation, legislation, supervision, policing of fraud etc.
All the things which are absolutely necessary
for the traditional financial institutions to function.
Instead bitcoin takes a truly and radically different approach.
It is an attempt to build a financial infrastructure
based on entirely new premises.
A sort of peer-to-peer financial anarchy run by people who trust no one.

\subsection{The Cryptographers' Dream}

The main proposition is something which we frequently see in cryptography.
We call it a cryptographer's dream:
a dream about the world which functions with participants which do not
see each other, do not trust each other a lot,
and yet are able to somewhat function
and achieve some sort of ``secure function'' or prevent fraud from being committed.
An attempt to build systems which remove the necessity of having
trusted parties such as financial institutions
and other businesses,
intermediaries, or providers of services.
Or at the very least, to greatly decrease the trust assumptions which are necessary.
However finance and payment is not an idealized cryptography world.
Here the possibility of building such systems is quite surprising, disturbing and needs yet to be demonstrated.

The current bitcoin ecosystem remains excessively fragile:
there is essentially one software distribution \cite{BitcoinMainSoftwareDistribution}
which implements a ``full network node'', and which is the critical
infrastructure necessary for bitcoin to function.
This software is certainly a critical point of failure \cite{blockchainfork}.
More importantly,
not all participants in the bitcoin network
are the same. 

\subsection{Critical Nodes: Miners}

In fact only very few of the full network nodes are so called miners.
These miners do two very important things which are very closely related:

\begin{enumerate}
\item
First of all they actively participate
in approving and verifying the correctness
of bitcoin transactions.
Ordinary network participants do not need to perform these checks
except maybe for the very few transactions
in which they are personally involved.
However miners are expected to check the correctness of transactions
and approve them. Collectively miners generate a consensus
which is a sort of official bitcoin history
(also known as {\bf the Main Chain}) which
allows to prevent fraud such as double spending.
\item
More importantly miners are people who manufacture (mint) the currency:
they spend a lot of computing power on solving a specific type of
cryptographic puzzles (cf. \cite{MiningUnreasonable})
and in exchange they cash rewards
in the form of freshly created coins.
They also have an additional source of income which comes
from the fees on all the transactions they have approved.
\end{enumerate}
The solutions to these cryptographic puzzles are made public
and are an essential part of the official public ledger
of all bitcoin transactions ever made.
These solutions are called {\bf bitcoin blocks}.
Miners spend money on purchasing equipment for bitcoin mining
and on electricity.
The equipment for bitcoin mining has with time become increasingly sophisticated
and nowadays mining is done
with highly specialized computing devices
known as ASIC miners
which exist only for this purpose, cf. \cite{MiningUnreasonable}.
This is clearly {\bf a violation} of the original idea of bitcoin by Satoshi Nakamoto,
who very clearly postulated that {\em each node} should be collecting recent transactions
and trying to create new blocks in the currency, cf. Section 5 of \cite{SatoshiPaper}.
However such specialized devices allow to create bitcoins
while using as little as 10,000 times less energy
than with ordinary PCs, cf. \cite{MiningUnreasonable}.

The two activities of the miners are very closely related.
The implicit assumption is that miners who approve transactions ``honestly''
or as expected by the designers of this digital currency
\cite{SatoshiPaper} will have in principle higher chances to reap the rewards
which are the precisely the incentive for miners to support the digital currency.
This is in principle, as in practice we are going to show
that sometimes miners will deviate from
the ideal behavior for profit,
and that there will not suffer from a penalty of any kind.
Moreover we will show that such behavior can remain totally invisible
and sometimes it simply cannot be detected, not even in theory.


It is possible to take a view that what bitcoin really is,
is just a (somewhat decentralized) high-tech business venture,
run by a group of people (miners) which make money from this activity.
As such miners have vested
interests and do not necessarily represent
the interest of the majority of network nodes.
Bitcoin is simply not (or not yet)
this sort of decentralized utopia network
which belongs to no one and works for everybody,
which sometimes it claims to be.
%
We need to stop claiming that bitcoin is an utopia which it never was,
and see it as a game with multiple participants with their interests.

\subsection{Big Uncertainties}

Bitcoin is an attempt to build a system
able to carry out simple financial transactions
such as payment for goods and services over large distances
(for example in the digital economy).
However this attempt is imperfect and fragile. 
The software specification of bitcoin
is not written in stone
and it is likely to evolve.
This possibly in very strange directions dictated
by minorities such as some miners.
Bitcoin cryptography is likely to be just broken in the long run.
The design could also be subverted.
The creator of bitcoin is an anonymous person or group
and no well-known authority in cryptography or information security
have certified that bitcoin is in some sense secure.

There is no such thing as a free self-governing space.
Sooner or later some sort of bitcoin governance must emerge.
In contrast with the bitcoin software developers
which are well-known public figures
and whom we tend to trust \cite{BitcoinMainSoftwareDistribution},
miners are an obscure group of anonymous people
organized in a handful of groups or pools such as
BTC Guild, ASIC Miner, GHash.IO, etc.. 
cf. \cite{BitcoinExistingPools,BitcoinPoolStrategiesAvoidCheating}.
In most cases we don't even know in which country they reside.
Moreover they tend to hide for tax evasion reasons:
billions of dollars of (potential or/and realized)
profits have been made
by miners in the recent years.
Most of these profits
remain invisible to tax authorities.
A great majority of bitcoin users use pseudonyms
and their real identity is unknown.

In this paper we study the question of miner strategies.
What miners can do to increase their already very substantial revenue.
In particular we focus on subversive strategies however
strange and farfetched they may appear.
We point out that miners are subjected to grat many moral hazards
which in the long run they might not be able to resist.
We also postulate that these questions should be discussed
openly if we want the security of digital currencies to improve.

\section{Short Description of How Bitcoin Works}

It would be controversial to say that there exists
such a thing as a unique and authoritative bitcoin specification.
This would be somewhat contrary to the spirit of bitcoin.
The original paper contains very few actual details and
as we have already seen it is not always followed
in the bitcoin practice.
However at this moment in history,
we have essentially
one dominant form of bitcoin software \cite{BitcoinMainSoftwareDistribution} and
the following web page claims to contain the primary official bitcoin specification \cite{BitcoinTechSpec}.
However the bitcoin specification is not written in stone.
In reality the bitcoin specification is in a constant flux and bitcoin
is officially claimed to be experimental rather than mature.
This software changes with time and in spite of being open source it is very obscure.
It uses scripts and rich in functionality libraries such as OpenSSL
which imply the possibility to easily extend their functionality in the future in maybe unexpected ways.
Other independent code bases are emerging, and in the future there is no guarantee that there will be an universal agreement on
what bitcoin is and what it should be. These questions could just be decided by consensus.

Various authors describe bitcoin to a certain extent and they usually omit many essential details.
There are many common misconceptions about bitcoin in the press: for example we frequently hear that
bitcoins are encrypted. In fact bitcoin can function entirely without any encryption.
It is all about integrity and authenticity of transactions.
These goals are achieved through digital signatures and cryptographic hash functions.

In one sense the whole of bitcoin is just
a sort of distributed electronic notary system which works by consensus
and with peer-to-peer payment as the main practical application.
Below we provide a short, concise 
description of how bitcoin works
for the purpose of this paper.

\begin{enumerate}
\item
We have a decentralized network of full bitcoin nodes which resembles a random graph.
Network nodes can join and leave the network at any moment.
\item
A public ledger of all transactions is maintained and
it is used to record payments with units called bitcoins.
\item
Bitcoins are divisible and are in great simplification
like digital objects
stored on the computers of the network participants.
\item
Initially, when bitcoins are created,
they are attributed to any network node
willing and able to spend sufficient computing power
on solving difficult cryptographic puzzles.
\item
It can be seen as a sort of lottery.
Currently 25 BTC are attributed to just one
winner every 10 minutes on average
(however this quantity decreases with time).

We ignore the question of
how exactly the difficulty is adjusted to achieve that
and related attacks, see
\cite{BlockDiscardingNetworkSuperiority}
\item
Miners repeatedly produce a double SHA-256 hash H2 of
a certain data structure called a block header
which is a combination of
events in the recent bitcoin history
and which process is described in details in \cite{MiningUnreasonable}.
\item
This H2 must be such that when written as an integer in binary
it will have some 64 or more leading zeros.
\item
More precisely,
in order to produce a winning block,
the miner has to generate a block header such that
its double SHA-256 hash H2
is smaller than a certain number called \mbox{target}.
\item
This can be seen as essentially a repeated experiment where H2 is chosen at random.
The chances of winning in the lottery are very small
and proportional to one's computing power multiplied by $2^{-64}$.
This probability decreases with time as more miners join the network
and the difficulty to find one block increases.
\item
If several miners complete the winning computation only one of them
will be a winner which is decided later by a sort of majority vote.
This problematic
situation is called a fork and currently happens less than 1 $\%$ of the time,
see Table \ref{PercWastedOverTime} on page \pageref{PercWastedOverTime}.
\item
Existing portions of the currency are defined {\bf either} as outputs of a block
mining event (creation)
{\bf or} as outputs of past transactions (redistribution of bitcoins).
\item
The ownership of any portion of the currency is achieved through digital signatures.
\item
Each existing quantity of bitcoin identifies its owner by specifying his public key or its hash.
\item
Only the owner of the corresponding private key
has the power to transfer
this given quantity of bitcoins to other participants.
\item
Coins are divisible and transactions are multi-input and multi-output.
\item
Each transaction mixes several existing quantities of bitcoins
and re-distributes the sum of these quantities of bitcoin
to several recipients in an arbitrary way.
\item
The difference between the sum inputs and the sum of all
output amounts is the transaction fee.
It belongs to one winning miner
who have managed to include this transaction is his block.
\item
Each transaction is approved by all the owners of each
input quantity of bitcoins
with a separate digital signature approving the transfer
of these moneys to the new owners.
\item
The correctness of these digital signatures is checked by miners.
\item
Exactly one miner approves each transaction.
However blocks form a chain and other miners will later approve this block,
and at this moment they should also check all the signatures.
\item
All this however is effective only for blocks which are in the
dominating branch of bitcoin history (a.k.a. the Main Chain).
Until now great majority of events in the bitcoin history made it to become the part of this official history.
\item
In theory every bitcoin transaction could later be invalidated.
The common solution to this problem is to wait for a small multiple of 10 minutes
and hope that nobody will spend additional effort just in order to invalidate one transaction.
Moreover even if there is a fork, we can hope that everyone was honest,
and our transaction will be included in both versions of the history.
\item
In practice the problematic wasted effort
which does not become the part of the official history and could lead to
invalidation of past transactions remains marginal.
It represents less than 1 $\%$ of the total computational effort,
see Table \ref{PercWastedOverTime}.
This is because the propagation in the bitcoin network is quite fast:
the median time until a node receives a block is 6.5 seconds whereas the average time is 12.6 seconds,
see \cite{ForksPropagationDecker,ForksPropagationDecker2}.
However
after 40 seconds there still are $5\%$
of nodes that have not yet received the block.
\item
Overall the network is expected to police itself,
miners not following the protocol risk that their blocks
will be later rejected by the majority of other network participants.
Such miners would simply not get the reward for which they work.
\end{enumerate}

\section{How Pools Work}

In this section we describe very briefly how pools work and define the concept of a ``share''.

\begin{enumerate}
\item
The main reason why miners mine in pools is the reduction of the uncertainty:
people want to have regular income and they do not want to play the lottery.
\item
Miners in one pool mine with the public key of the pool manager
which cashes the gains and re-distributes
them according to their contribution in the join effort.
\item
If miners find a block which hashes to a value which starts with many zeros,
for example 32 zeros,
they send such value to the pool manager.
\item
This is a proof of effort
worth $2^{32}$ attempts of computing the double hash of Thm. \ref{HashSpeed1.86}
and which can be easily checked by the pool manager.
\item
We call such a contribution {\bf a share}.
\item
Miners send their shares to the pool manager as soon as they find them.
\item
The process of computing a share is the same process as the process of computing a valid block
however the difficulty level is lower.
Some of these shares have a hash which has more than 32 zeros, for example 64 or more zeros.
\item
Such winning shares are with high probability computed by just one miner,
cf. Table \ref{PercWastedOverTime}, and with high probability
they will form a new valid block for the bitcoin network.
\item
A winning share allows the pool manager
obtain approximately 25 BTC at the present moment,
cf. \cite{MiningUnreasonable} for more details.
This money is distributed among the pool participants.
\end{enumerate}



\section{On Statistical Approximation of the Bitcoin Mining Process}

In this section we assume that a mining pool is a
static union of miners who have put together
their computing power and share rewards more or less uniformly,
cf. \cite{BitcoinExistingPools,BitcoinPoolStrategiesAvoidCheating,RosenfeldPoolRewardPaper,RosenfeldPoolRewardMethods}.
We basically consider an arbitrary fixed subset of bitcoin mining devices.

\subsection{On Repeated Events}

Bitcoin mining is based on hash functions the output of which is expected to behave essentially
as independent random variables.
Repeated events are governed by the laws of statistics such as the Law of Large Numbers
and more precisely we have the following well-known result:

\begin{thm}[Central Limit Theorem]
\label{CentralLimit}
Let $\{X_1,\ldots,X_n\}$ be a sequence of independent and identically distributed random variables
each of which having the expected value $\mu$ and finite variance $\sigma^2$.
Then
$$
\lim_{n\to \infty}
Pr(\frac{
\sum\limits_{i=1}^n X_i - n \mu}
{\sqrt{n} \cdot \sigma}
\leq z ) = \Phi(z),
$$
where $\Phi(z)$ is the probability that
a standard normal variable is less than $z$.
\end{thm}

Informally and for the purpose of this paper,
for any event which is repeated many times,
whatever is the actual probability distribution, if we repeat
the experiment $n$ times for a large $n$ and add the outcomes $X_i$,
the resulting variable has a standard deviation of approx. $\sqrt{n}$ times
the original standard deviation $\sigma$.
Moreover any larger or smaller deviation will be subject to the Gauss error function
and large deviations will be extremely unlikely,
for example the probability to be outside of 8 standard deviations
is as small as $2^{-50}$, cf. \cite{wikistdev}.

In Theorem \ref{CentralLimit} the number
of events $n$ is assumed to be very large.
However it can be also applied (indirectly)
to the analysis of rare events such as bitcoin mining events
which can be shown to be a by-product of
a very large number of more basic events.

\subsection{On Rare Events and Bitcoin Mining}
\label{BitcoinMiningLawSection}

It is easy to show that the following result holds approximately
(but not exactly!) for any group of mining devices
and therefore also for any fixed
group of miners and therefore for
mining pools with a static set of members.

\begin{thm}[The Law Of Bitcoin Mining]
\label{BitcoinMiningLaw}
If in any given period of time a group
of miners has the computing power
which allows it to mine $K$ blocks in expectation,
then the standard deviation of the number
it will actually mine
will be approximately $\sqrt{K}$,
following the Poisson distribution.
Moreover following Thm. \ref{CentralLimit} the
probability that we will be at a certain number of
standard deviations will be governed by the Gauss error function.
\end{thm}

\noindent\emph{Proof 1:}
In order to obtain this result we need to assume that at any given moment in history, the probability
for a pool to mine the block is constant, and it does not depend on previous events (no memory).
This is only an approximation as pool members leave and join, the difficulty to mine a block is variable,
and also the hash power of the competition in the bitcoin network fluctuates slightly.
Then we have the classical case of a Poisson process
in which the mean is equal to its variance
\footnote{This distribution is sometimes called the {\em Law of Small Numbers}
from a title of an old classical book \cite{Bortkiewicz}
which has popularized the Poisson distribution
as a tool which models quite
well various sorts of real-life events. 
},
and therefore the standard deviation is the square root of the mean which is $\sqrt{K}$,
cf. \cite{wikiPoisson,RosenfeldPoolRewardPaper,RosenfeldPoolRewardMethods,Bortkiewicz}.

\noindent\emph{Proof 2:}
Here is an independent derivation and a specific example rooted in the bitcoin mining.
It is made to be as realistic as possible and very close to what
is actually happening in the bitcoin network.
It is the based on the well-known property that the Binomial distribution converges toward the Poisson distribution.

We consider that there are many miners trying to mine a block in a mining pool.
Miners of variable size will be modeled
as combinations of several miners $X_i$ of equal size,
which makes that $n$ is quite large.
We have $X_i=1$ if a given miner
(or a unit of all available computing power)
finds the winning block.
Most of the time $X_i=0$.
More precisely each miner will be modeled
as a combination of mini miners
each of which evaluates the double compression function of
Theorem \ref{HashSpeed1.86} just once in one unit of time
(here we ignore the cost of this process and possible speed-ups).
Let assume that by definition the winning block computes a
hash which has exactly 64 zeros,
which is close to the reality. 
Thus in approximation,
in a pool which has produced say $K$ blocks in a period of time,
there were $n=K\cdot 2^{64}$ trials.
We have $n$ independent random variables $X_i$ with mean of $\mu = 2^{-64}$ each.
They $X_i$ are equal to one if and only if
one of the hashes tried has succeeded
to produce a winning block with 64 zeros.
It is straightforward to compute the variance of each of these variables:

\vskip-6pt
\vskip-6pt
$$\sigma^2 = (1-\mu)^2 \mu+(0-\mu)^2 (1-\mu)=\mu-\mu^2\approx \mu.$$
\vskip-2pt

Now for
$X_1 + \ldots + X_n$
and since all the trials are independent
the variance will be $n\sigma^2\approx n\mu=K$.
Then however the standard deviation is equal
to its square root which is $\sqrt{K}$.

{\bf Remark 1:}
Needless to say, in the real life this analysis is just an approximation,
valid if the network power is distributed uniformly in time and if the difficulty if constant.

{\bf Remark 2:}
When we apply Theorem \ref{BitcoinMiningLaw} in practice
we frequently need to make {\bf another approximation}.
We do not know what exactly was the computing power of one pool, we can just estimate it
from the number of blocks mined.


\section{Basic Attack Strategies For Subversive Miners}

The research in the area of miner strategies have evolved from
simple attacks with large majority
towards less obvious attacks
which operate in more discreet ways.
It this paper we describe a number of attacks
in which the basic strategic objective
is to assure some group of malicious miners
a certain unfair advantage,
for example to obtain an expected revenue from mining
higher than their fair share
based on the contributed computing power.

\subsection{The Pool Hopping Attack}
\label{PoolHoppingAttack}

The ``Pool Hopping Attack'' is described in \cite{RosenfeldPoolRewardPaper,RosenfeldPoolRewardMethods}.
The main idea in this attack is that if a miner mines in a pool in which a lot of shares have already been submitted
and no block has yet been found,
he will gain less in expectation because the reward will be shared with the miners who have contributed to this pool.
Therefore at a certain moment it may be profitable to stop mining in this pool and contribute elsewhere.
This remains valid even if the pools penalize leavers and refuse to pay for their contribution
if they do not mine for a complete ``shift''.
It is still profitable for miners to quit and mine for another pool (or mine independently).

This attack works more or less well depending
on how exactly pools are managed and also depending on the actions of other miners,
cf. \cite{RosenfeldPoolRewardPaper,RosenfeldPoolRewardMethods}.
It can be shown that hoppers will earn more than normal ``continuous'' miners.
Various reward and pool management methods have been proposed in order to
discourage pool hopping 
and some reward methods can be shown to be immune to this attack,
we refer to \cite{RosenfeldPoolRewardPaper,RosenfeldPoolRewardMethods}
for more details.

\medskip

In this paper we assume that miners
do not change pools frequently 
and that in the long run,
whatever is the reward method,
they are paid in approximately in
proportion to their contribution.

\subsection{The Mining Cartel Attack}
\label{CartelAttack}

The ``Mining Cartel Attack'' is described in \cite{CartelAttack}.
It is an attack in which a large fraction of miners such as $50 \%$
decide to ignore some or all blocks generated by miners
which are not members of the cartel.
This allows dishonest miners to achieve higher gains.

\subsection{A Difficulty Raising Attack}
\label{DifficultyRaisingAttack}

In this recent attack a powerful attacker is
secretly preparing an alternative version of the blockchain.
At the same time he is manipulating the automatic difficulty
adjustment mechanism in his secret chain in order to increase the
probability of eventually that
his chain will be recognized as surpassing the public honest chain.
If this happens, the attacker reveals his secret chain.
This can be used to commit double-spending. 
See \cite{BlockDiscardingNetworkSuperiority} for more details.


\section{Confidential Cryptographic Optimization Attack}
\label{CryptoOptimizationAttack}

This is a new attack and is based on the
idea that miners can improve and optimize the mining process.
We refer to \cite{MiningUnreasonable}
for the historical evolution of bitcoin mining
and how the power consumption of bitcoin mining
have evolved
due to the technology shift from all purpose
to specialized silicon devices which are designed and made
just for the purpose of bitcoin mining.
If we assume that same silicon technology is achievable for
all miners sooner or later, there is also a space for algorithmic improvements which
will be an additional source of competitive advantage.
A recent paper shows that the mining
process can be improved by some 38 $\%$ by
cryptographic optimization \cite{MiningUnreasonable}.
This is not negligible knowing that
literally megawatt hours of energy
are 
spent on bitcoin mining
\cite{MiningMegaWattsEnvirDisaster}.
Here is the main result:

\begin{thm}[Cryptographic Optimization of Bitcoin Mining]
\label{HashSpeed1.86}
The amortized average cost of trying
one set of data to see if the double hash of it
with SHA-256
has 64 or more
leading zeros is only at most about
1.86 computations
of the compression function
of SHA-256 instead of 3.0 in naive version.
This represents an improvement by $38 \%$.
\end{thm}
\vskip-5pt

\begin{figure}[!here]
\centering
\begin{center}
\vskip1pt
\vskip1pt
\includegraphics*[width=3.5in,height=3.2in,bb=0pt 0pt 1158pt 666pt]{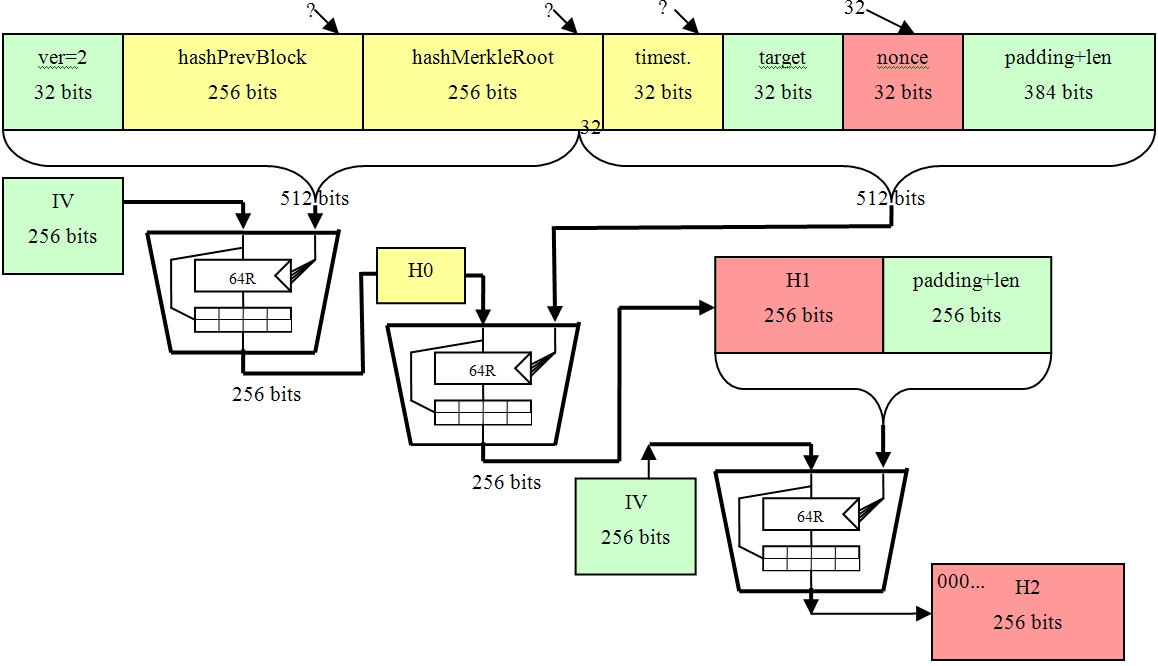}
\vskip-0pt
\vskip-0pt
\end{center}
\caption{
The process of bitcoin mining according to \cite{MiningUnreasonable}.
}
\label{BitcoinCISOProblem2}
\vskip-3pt
\vskip-3pt
\end{figure}

\noindent\emph{Proof:}
We decompose the double application of SHA-256
which is the essence
of bitcoin mining into 3 applications
of the SHA-256 compression function
(cf. Fig. \ref{BitcoinCISOProblem2})
in which many
individual pieces of data can be pre-computed and are essentially constants
over many computations, or change in an incremental easy to predict way.
Moreover some of the final computations can also be omitted
as the result is incorrect most of the time and we can have early rejection with incomplete data.
A complete proof is given in \cite{MiningUnreasonable}.
%
%

\noindent\emph{Application:}
Is there a place for subversive strategies here? Yes.
There is only a handful of cryptologists who are able to improve
the core process of bitcoin mining.
Paying these people to develop confidential improvements
which will benefit only one group of miners,
not the majority of miners is certainly
an obvious subversive strategy.
However
since the publication of \cite{MiningUnreasonable}
the playing ground has been somewhat leveled.
It is possible to believe the figure of 1.86 is close to the theoretical limits
and that further gains from this strategy
can no longer be very substantial, maybe just 1 or 2 $\%$.
This is not yet the most disturbing subversive strategy for miners we can think of.

\newpage
\section{Selfish Mining Attack}

At the end of 2013 almost
all the newspapers have written
about one particular paper about bitcoin by
Cornell researchers Eyal and Sirer
\cite{BitcoinBrokenGrowingSelfishPoolStrategy}
with a title:
\emph{Majority is not Enough: Bitcoin Mining is Vulnerable}.
This paper
is basically a game theory paper which shows that
in the current bitcoin currency systems miners have a
particular sort of subversive strategy,
a version of which was in fact
also independently invented and studied
by Bahack cf. Section \ref{BlockDiscardingAttack} below.

This strategy is called {\em Selfish Mining} in
\cite{BitcoinBrokenGrowingSelfishPoolStrategy}
and it is also studied as strategy $st_1$ which is one
of {\em Block Discarding Attack(s)} studied in
\cite{BlockDiscardingNetworkSuperiority}.  
The attack consists of information concealing
in a very selective and special way,
and revealing it just in time, also in a selective way.
It is very much counter-intuitive
as it seems that bitcoin is based on self-interest,
and that the first interest of the miner is to have his
part of the effort disclosed as soon as possible,
in order to be used by other miners in the mining community.
This is because only then the efforts of the miner
will be paid for: the miner obtains bitcoins and will be able to
pay for his hardware and electricity expenses spent on solving the cryptographic puzzles.
However in the subversive strategy miners are trying to confuse other miners
to also waste their efforts and this in higher proportion than themselves.

It is a surprising and highly technical proposition which nobody yet understands fully.
It appears that there exists a strategy for miners in which miners
sometimes voluntarily do not publish their results as soon as they find them,
and rather delay this publication
by which process they waste some of their computational effort and put themselves at risk
of not getting the reward.
However at they same time they also create some confusion in the network
which makes other miners to waste even more effort (in proportion).
It is entirely about making other miners to try to mine on blocks such that
their effort is likely to be lost, because another branch of the bitcoin network
exists and is kept confidential.
This for a relatively short time, in order to minimize losses and maximize the gain.
This is quite surprising and counter-intuitive.
However it is shown in \cite{BitcoinBrokenGrowingSelfishPoolStrategy}
that subversive miners can obtain rewards which
are in proportion (maybe only slightly) bigger
that their share of computing power.

Here is a simplified (incomplete) description
of the selfish attack.
Our intention is to show
how this attack looks like in practice,
what are the guiding assumptions which command it, 
and to see which events are expected to be the most frequent. 
\vskip-1pt
\vskip-1pt
\begin{enumerate}
\item
{\bf Assumption 1}.
Most of the time subversive miners behave like normal miners.
If there is a single longest chain in the public
blockchain, all miners try to extend it.
In absence of a fork, and in absence of ``secret'' blocks which
some miners keep for themselves,
the most likely winner block is the same for everyone.
This ``consensus'' situation will be very frequent.
Here subversive miners have no particular advantage.
\item
If the selfish miners find a new block (B1) first,
they do not publish it instantly,
and share it only inside the selfish pool.
We say that selfish miners now lead by 1 step.
\item
The number of possible situations is greatly reduced by the following assumption:
{\bf Assumption 2}:
At any moment during the attack there are up to two competitive public branches
one of which can have a secret extension.
We have either just one branch
(with possibly a secret extension by the attackers)
or a public fork with two branches of equal depth $k\geq 0$.
In the case of a fork one branch is composed solely of honest miner's blocks
and the other is composed solely of attacker's blocks
which at moments can have a secret extension.

On page 4 of \cite{BlockDiscardingNetworkSuperiority} it is claimed
that this assumption can be made for all ``interesting'' strategies
for the subversive miners and that other strategies are sub-optimal.
\item
Concealing a block (or more) leads to different views.
Honest miners will be mining 1 (and rarely more) steps behind the selfish miners.
The effort of honest miners during this time is likely to be wasted,
we call this property (R1).
\item
The effort of subversive miners could also be wasted
if secret block(s) are kept secret for too long.
There is a risk is that some secret block(s) would never be used.
We call this property (R2).
\item
For this reason the block is kept secret only for a short time.

Further detailed rules of the attack have been designed in such a way in order
to maximize the gain from (R1) and minimize the losses from (R2).
\item
We assume that the subversive miners lead by 1 block (B1).
Most of the time honest miners will also find an equivalent
block and destroy the lead of 1 block. We call such events (C1).
Sometimes the lead increases which we call (C2).
\item
When (C1) the subversive miners instantly reveal their secret block and create a public fork.
Now the subversive miners do not lead, however they have started mining on their secret block (B1) earlier.
\item
During a fork situation selfish miners mine on their branch (B1),
while honest miners mine on both branches depending on the network propagation.

Interestingly the selfish miners want other honest miners to also mine on the same block (B1):
this block costed a lot to produce and
the risk of not being paid for this block is very high (due to the fork). 
This is the main event when subversive miners could lose from (R2).
\item
Going back to the initial situation before a fork with 1 secret block,
we have the less frequent possibility (C2) that subversive miners
extend their secret chain by yet another block (leading now by 2 or more blocks).
Then they keep mining on their secret branch as long as the lead is 2 or more and they publish
some blocks from the secret branch each time the honest miners find a new block on their branch.
The honest miners may then or not switch to their branch and they are permanently exposed to (R1).
However subversive miners avoid making the lead become 1.
This happens however with high probability due to the actions of the honest miners.
Then the selfish miners publish all their blocks in order to convince the honest miners
to definitely switch to their branch due to Assumption 1 above
and achieve the initial ``consensus'' state.
We should note that 
situations with lead 2 or higher are not very frequent.
\end{enumerate}
\vskip-5pt

Overall the strategy is very complex and it is not obvious to see if and when it is profitable,
see \cite{BitcoinBrokenGrowingSelfishPoolStrategy}.

\subsection{Discussion of the Selfish Mining Attack}

Assuming that the main result of \cite{BitcoinBrokenGrowingSelfishPoolStrategy} is correct
(it was confirmed by computer simulations)
we are at present moment
{\bf not at all convinced by the claims of this paper}.
On the contrary we believe that the claims of this paper ar exaggerated and only valid
in a certain context.
The reasons for this are:

\begin{enumerate}
\item
One cannot claim that ``Bitcoin is broken'' just because some subversive strategy exists.
Other even better strategies with higher expected gain may exist.

\item
The paper \cite{BitcoinBrokenGrowingSelfishPoolStrategy}
makes one very explicit assumption which potentially invalidates the
overall result. On page 5 we read:
\emph{
For simplicity, and without loss of generality,
we assume that miners are divided into two groups,
a colluding minority pool that follows the selfish mining strategy,
and a majority that follows the honest mining strategy (others).}
{\bf However there is no reason to assume that if there
is a subversive strategy,
there will be only one group following it.}
On the contrary. Because the current proposal
is about secrecy (concealing information which should not leak outside of one group),
it is very strange to assume that there will be only one such group.
It is not clear at all that the effect of several such groups would be the same
or that such groups will have incentives to merge into one group.
Several competing subversive groups of equal size might be competing against each other
and this will probably invalidate or decrease the benefits of each other's strategy.

\item
For large pools it may become infeasible to keep secrets from others.
In \cite{BlockDiscardingNetworkSuperiority} it is claimed that such an attack
is rather applicable to solo miners. 

\item
There is no evidence that miners actually mine on the first block they receive as claimed in \cite{BitcoinBrokenGrowingSelfishPoolStrategy}.
Though the bitcoin source code is open \cite{BitcoinMainSoftwareDistribution},
the code of miners is nowadays in great majority closed and we simply
don't know how advanced it is and what improvements or optimizations (cf. Thm. \ref{HashSpeed1.86})
have already been made.

\item
Each strategy has adoption thresholds. 
Gains need to be demonstrated to be substantial
or they will not make any difference.
It is simply not correct to assume that miners
will switch to a certain strategy or join a certain pool
if the benefits are just 0.01 $\%$ in theoretical mathematical expectation of  gain.
This knowing that gains are not evenly distributed, have a large variance and
therefore small benefits
are just not visible to small participants.
It will be in fact difficult to convince people that gains are real.
Larger participants may of course optimize
their operations with utmost scientific precision and care,
however one of the claims of the paper \cite{BitcoinBrokenGrowingSelfishPoolStrategy}
is that even very small groups of participants
(without a lower limit on their market share)
would be interested in adopting a subversive strategy
or/and would be compelled to join a growing group of ``selfish miners''.
This is simply not true.
\item
Even if miners would in theory join some sort of subversive pool
as claimed in \cite{BitcoinBrokenGrowingSelfishPoolStrategy},
they could deviate from the expected ``honest'' behavior inside that pool.
For example members of the pool can apply an attack
we describe later in Section \ref{BlockWitholdAttack}
in which a sub-group of miners works against the pool and all the other miners
and achieve higher gain than others.
\item
There is another technical argument
which will be presented later
when we are going to comment on
the historical data in Table \ref{PercWastedOverTime}
and view the possible advantage to be gained
as possibly only a small proportion of
an already small proportion
of ``wasted'' computational effort.
\end{enumerate}

In this paper we do not attempt to solve this problem of what is the best
strategy for bitcoin mining neither we claim that this problem is solvable.
We just point out that the problem of selfish behavior is immaterial as of now.
We have analysed the historical data of blocks which have been wasted in the bitcoin network
for the whole history of this network.
In Table \ref{PercWastedOverTime} we present the results
knowing that the timing of the bitcoin network is one block published every 10 minutes.
The table covers the whole bitcoin history since the system
has started to function in early 2009.


\vskip-0pt
\vskip-0pt
\begin{table}[h!]
  \caption{Percentage of blocks ever mined wasted due to a fork
and wasted blocks which are children of wasted blocks}
    \label{PercWastedOverTime}
\vskip-0pt
\vskip-0pt
$$
\begin{array}{|c|c|c|c|}
\hline
\mbox{blocks}   
& wasted & child(wasted)\\
\hline
\mbox{less than~}140,000   
 & 0.00 $\%$ & 0.00$\%$ \\
\hline
\mbox{140,000-149,999}   
 & 0.21 $\%$& 0.00$\%$ \\
\hline
\mbox{150,000-159,999}   
 & 0.27 $\%$& 0.01$\%$ \\
\hline
\mbox{160,000-169,999}   
 & 1.01 $\%$& 0.01$\%$ \\
\hline
\mbox{170,000-179,999}   
 & 1.77 $\%$& 0.29$\%$ \\
\hline
\mbox{180,000-189,999}   
 & 1.71 $\%$& 0.01$\%$ \\
\hline
\mbox{190,000-199,999}   
 & 1.15 $\%$& 0.01$\%$ \\
\hline
\mbox{200,000-209,999}   
 & 0.88 $\%$& 0.00$\%$ \\
\hline
\mbox{210,000-219,999}   
 & 1.05 $\%$& 0.00$\%$ \\
\hline
\mbox{220,000-229,999}   
 & 1.28 $\%$& 0.42$\%$ \\
\hline
\mbox{230,000-239,999}   
 & 0.78 $\%$& 0.00$\%$ \\
\hline
\mbox{240,000-249,999}   
 & 0.43 $\%$& 0.00$\%$ \\
\hline
\mbox{250,000-259,999}   
 & 0.67 $\%$& 0.01$\%$ \\
\hline
\mbox{260,000-now}   
 & 0.91 $\%$& 0.01$\%$ \\
\hline
\end{array}
$$
\vskip-1pt
\vskip-1pt
\end{table}
\vskip-0pt

In this table we show that the wasted computational
effort in bitcoin has been marginal so far,
and has even decreased with time
(mostly due to the work of Swiss researchers, see \cite{ForksPropagationDecker,ForksPropagationDecker2}).
Moreover it is possible to see that
the space for the gain from the subversive strategy
{\bf is even much smaller}.
Let us explain this in detail.
The only benefit we expect from the ``selfish strategy''
of \cite{BitcoinBrokenGrowingSelfishPoolStrategy}
is that subversive miners
will make other miners mine on blocks which later
become wasted blocks.
More specifically, there is only one place where the ``selfish strategy''
makes the honest majority of miners lose something in proportion higher
than for the subversive ``selfish'' miners themselves.
This is when the ordinary miners are going to waste
some of their computational effort on producing what
technically are children of wasted blocks in the directed graph
of all blocks ever mined.
{\bf This is currently only 0.01 $\%$ of miners' combined effort (!)}.

In appearance this just confirms what the authors of
\cite{BitcoinBrokenGrowingSelfishPoolStrategy} have written:
\emph{"To the best of our knowledge, [...] pools } (by which they mean miners)
\emph{[...] have been benign and followed the protocol."}
However we should think about it more than once.
It is obvious that,
if the column 3 of our table
has to increase substantially in the future
due to the adoption of some subversive mining strategies,
the column 2 also needs to increase in a very substantial way (!).
Then however, miners will simply not accept to waste a lot of computational effort
and real money spent on electricity.
It appears that Megawatts are currently spent on bitcoin mining,
see \cite{MiningMegaWattsEnvirDisaster,MiningUnreasonable}.
Miners will lobby to somewhat change the specification of bitcoin
in such a way that subversive strategies which make everybody
lose a large part of their effort are eliminated,
or at least that they remain as marginal as they currently are.
For example bitcoin might agree to 
penalize forks
which will decrease the benefits of subversive strategies which create more forks than usual,
such a countermeasure has been proposed in \cite{BlockDiscardingNetworkSuperiority}. 

To summarize we believe that the claims of \cite{BitcoinBrokenGrowingSelfishPoolStrategy}
are vastly exaggerated and that
it is rather a purely academic game theory result
of little practical importance.
Selfish mining is probably not the most practical
of the subversive miner strategies which we cover
in the present paper. 
In one sense the paper \cite{BitcoinBrokenGrowingSelfishPoolStrategy} is however very important:
it 
gives an explicit recommandation on {\bf how} bitcoin miners should
in the future select their blocks on which they mine in a presence
of a fork. They should do it at random from
all existing candidates and they should not trust
the timing of these blocks.
This recommandation of \cite{BitcoinBrokenGrowingSelfishPoolStrategy}
allows to eliminate the selfish miner strategies in practice.
It is also expected to work against some other subversive strategies
such as the recent Block Discarding Attack of \cite{BlockDiscardingNetworkSuperiority}.
As such it is likely
to be implemented by bitcoin miners very soon.

\section{A Block Discarding Attack vs. Selfish Mining}
\label{BlockDiscardingAttack}

The {\em Block Discarding Attack} is proposed in \cite{BlockDiscardingNetworkSuperiority}.
In fact the paper \cite{BlockDiscardingNetworkSuperiority} describes
and studies a number of more or less general
subversive strategies,
and the Selfish Mining attack of
\cite{BitcoinBrokenGrowingSelfishPoolStrategy}
can be seen as the simplest and the most basic block discarding attack
which is called st$_1$ in \cite{BlockDiscardingNetworkSuperiority}.

There is a number of differences in the
approach, assumptions, analysis and claims in both papers
(cf. Section 6.1. in \cite{BlockDiscardingNetworkSuperiority}).
The original Selfish miners attacks is designed for pools
while the paper \cite{BlockDiscardingNetworkSuperiority} explicitly
discards this idea claiming
that it is very hard for pools to keep their blocks secret
and that the attack is meant to be executed by solo miners.
The Block Discarding Attack described in
\cite{BlockDiscardingNetworkSuperiority}
have a variable called $ns$
which stands for ``Network Superiority''.
%
In \cite{BitcoinBrokenGrowingSelfishPoolStrategy}
they have a very similar concept: a propagation factor $\gamma$.
A close examination shows that
both concepts are the same.
In case of a fork with two equivalent blocks
$A$ and $B$,
the subversive miners
are trying to influence the honest miners
to mine on their branch $A$ and not on $B$.
The number $\gamma$ or $ns$ measures
the probability that honest miners mine on block A.


Another difference between two papers is that
in \cite{BlockDiscardingNetworkSuperiority} it is not clear how this influence
on the honest miners could possibly be achieved.
For example the attacker could be somewhat able manipulate the network latency
(or maybe control more ordinary network nodes which are not miners).
In \cite{BitcoinBrokenGrowingSelfishPoolStrategy} it is postulated
explicitly that this ``capacity to influence'' can happen
in a more ``natural'' way
because miners will mine on the block
which they have received first in the network.


\subsection{Comparison of Countermeasures}
\label{BlockDiscardingAttackCountermeasures}

The two papers also differ by recommendations on how to avoid the attack
and on profitability when the countermeasures are used.
Following page 6 of \cite{BlockDiscardingNetworkSuperiority}
we learn that the $st_1$ strategy
will be profitable if and only if 
the fraction of subversive miners is at least
$\frac{1-ns}{3-2ns}$.
The same formula should therefore apply to
the selfish attack of \cite{BitcoinBrokenGrowingSelfishPoolStrategy}.
The countermeasure suggested in \cite{BitcoinBrokenGrowingSelfishPoolStrategy}
is to mine at a random branch in presence of a fork.
This is equivalent to $ns=1/2$.
Then following the formula above
the attack is profitable if an only if
the attacker has at least 25 $\%$ of the total hash power.
On the other hand,
the fork punishment technique suggested in \cite{BlockDiscardingNetworkSuperiority}
does not try to change the $ns$ or $\gamma$
but simply makes certain blocks be rewarded less bitcoins.
This countermeasure can be claimed to be better and
achieve a higher profitability threshold than 25 $\%$,
thus making the selfish mining or similar attacks
less likely to be executed.
With fork punishment, it is possible to see that
the profitability threshold is such that
the attack is profitable only if the attacker has nearly 50 $\%$ of the total hash power.

\newpage
\section{A Block Withholding Attack}
\label{BlockWitholdAttack}

In this paper we study yet another subversive strategy for miners.
We believe that it is at least as realistic
as any other subversive strategy proposed so far. 
A block withholding attack was initially proposed by Rosenfeld in 2011, 
\cite{RosenfeldPoolRewardPaper}. 
Later in 2013, in Section 7 entitled Related Work
of \cite{BitcoinBrokenGrowingSelfishPoolStrategy} we read: 
\emph{
In a block withholding attack,
a pool member decreases the pool revenue by
never publishing blocks it finds.
}
In the earlier paper \cite{RosenfeldPoolRewardPaper}
two distinct  block withholding attack scenarios are described in Section 6.2.  
They are called ``Sabotage'' and ``Lie in wait''. 
However the first attack does not give the rogue miner any gain,
it just makes everybody loose. As for the second attack,
it is a complex block concealing attack similar
to the Selfish miner strategy of \cite{BitcoinBrokenGrowingSelfishPoolStrategy}
which we have criticized in this paper as possibly being not very realistic or not very practical. 
It should also be noted that block withholding attacks should not be confused 
with the block discarding attack of \cite{BlockDiscardingNetworkSuperiority}, 
cf. also Section \ref{BlockDiscardingAttack}. 

In what follows we are going describe a new sort of block withholding attack
which generalizes the ``Sabotage'' attack of \cite{RosenfeldPoolRewardPaper}.
We show that it is possible for rogue miners to profit from such an attack
which was not the case with the original attack.
We describe a concrete practical instantiation and
a concrete numerical example of a block withholding attack together
with an analysis of possible variants.
The basic version works as follows:

\vskip-3pt
\vskip-3pt
\begin{enumerate}
\item
We assume that all miners mine in pools, small and large.
Miners in one pool mine with the public key of the pool manager which later re-distributes the gains.
\item
We assume that the fees are very low.
Moreover we assume that miners
do not change pools frequently (cf. Pool Hopping attack \cite{RosenfeldPoolRewardPaper,RosenfeldPoolRewardMethods})
and that in the long run, whatever is the reward method,
they are paid in proportion to their contribution.
This is an approximation, see \cite{BitcoinPoolStrategiesAvoidCheating,BitcoinExistingPools}
for an overview of reward policies of some existing pools.
\item
We assume that there is a group of say $\alpha = 0.2 = 20 \%$
of rogue
miners.
We count 20 $\%$ in percentage of the total computing power of the network
measured in hashes per second.
\item
The rogue miners split in two groups of miners worth $\alpha/2 = 10 \%$ each.
Later in Section \ref{BlockWitholdAttackAlternativeVersions} we show that splitting in half is the optimal choice.
\item
Half of the rogue miners
representing $\alpha/2$ of the total computing power
participate in all the other existing pools in the
bitcoin community.
These ``infiltrated pools'' jointly command the total computing power of $1-\alpha/2$
including $\alpha/2$ of rogue miners.
This is done at random and in a distributed way,
uniformly in proportion
to the respective computing powers of these pools,
and under changing identities,
so that any subversive activity remains unnoticed.
\item
At the same time the rogue miners deploy their other $\alpha/2$
in their own minority pools which they control and dominate
(for example they create new small pools which other miners are unable to join in time).
\item
In each pool miners mine blocks for the pool manager
with the public key of the pool manager who controls the payouts.
\item
We assume that the pool managers are perfectly neutral and do not try to detect or prevent any unusual behavior.
\item
We assume that in the static case and in the long run,
various pool miner reward methods \cite{RosenfeldPoolRewardPaper,RosenfeldPoolRewardMethods}
have essentially the same effect:
they reward miners essentially in proportion to their efforts.
We also neglect the pool fees and the timing of the payments, cf. \cite{BitcoinExistingPools}.
\item
If miners find a share which has a hash with 32 zeros
and which is an undeniable proof of their effort,
they obtain a partial cash reward from the pool manager.
\item
Miners always send their shares to the pool manager as soon as possible.
\item
If ``normal'' miners find a share with 64 zeros
which allows one to generate a whole new valid block
for the bitcoin network and which makes the pool win 25 BTC,
\cite{MiningUnreasonable}, they send it to the pool manager.
\item
Rogue miners behave differently: they {\bf drop it},
and do NOT report it to the pool manager.
They basically destroy (erase) this share
which is worth a lot of money to the the pool manager
but nearly nothing to them
(they cannot benefit from the 25 BTC which this block brings because
they do not know the private key known to the pool manager).
\item
These events are very rare and happen 0 times for the majority of honest miners,
therefore the pool managers cannot see if these blocks are not reported.
In larger pools pool managers might discover that over the time they earn less
money that expected from their effort expended in the network,
cf. Section \ref{HaveRogueBeenDone},
however they will never be able to know which miners have cheated.
\item
Rogue miners are nevertheless paid for all of their efforts
(the difference 
is negligible).
\item
In our attack miners get paid for imitating honest miners
and sharing the reward obtained by others.
At the same time they mine for their own account
with half of their 20 $\%$ capacity.
Here they behave normally and do not waste any results.
\item
Overall it is easy to see that the rogue miners get
EXACTLY the same monetary return
on their computing power contributed in public pools they have joined,
and a strictly higher return in their private new pools.
\item
In the ``infiltrated pools'' the
shares contributed are equivalent to the actual
computing power of $1-\alpha/2$ however the winning blocks are
only generated by the honest majority of $1-\alpha$ in the hashing power.
All the miners in the ``infiltrated pools''
will see their monetary gains
uniformly reduced by a factor of
$\frac{1-\alpha}{1-\alpha/2}=\frac{80}{90}\approx 0.88$.

\item
Comparatively the other $\alpha/2=10 \%$ of rogue miners's capacity not involved in the well-known
``infiltrated'' pools do get a higher reward NOT reduced by the factor of $0.88$ which makes
it about
$1-\frac{1-\alpha/2}{1-\alpha}\approx $
13 $\%$ higher in proportion.
\item
Overall the rogue miners get paid a bigger share than other miners
and the difference is about $6\%$ to benefit the rogue miners.
This is because only half their mining capacity benefits from the
13 $\%$ higher returns, more generally the gain can be computed as:
\vskip-3pt
\vskip-3pt
$$
\frac{1}{2}\frac{1-\alpha/2}{1-\alpha} + \frac{1}{2} - 1 =
\frac{\alpha}{4(1-\alpha)}
$$
\end{enumerate}

With this attack the rogue miners obtain
a share of reward higher than their computing power contributed to the network
(which was also the main result of \cite{BitcoinBrokenGrowingSelfishPoolStrategy}).
It is easy to see that there is no easy way to stop this attack from happening.

\subsection{Can Block Withholding Attacks Be Prevented?}
\label{BlockWitholdAttackPrevent}

Our attack shows that pools can function well only if the pool manager can trust the pool participants.
We do not know if any of the current pools have implemented any countermeasure against this attack.
In \cite{BitcoinPoolStrategiesAvoidCheating} we read:
``withholding of good blocks by the clients is prevented by the server's possession of the private key''.
This is simply not true. It is clear that this does {\bf not} prevent the attack in the slightest.
It is more than probable that the only reason why miners have been benign
and honest so far was that they ignored the existence of various subversive strategies.

The only defense against this attack we can see is that pools should involve ONLY people
which we personally know and trust,
and the pool manager should simply dissolve and close a pool as soon
as he notices that it is earning less than expected from its computational effort.

\subsection{Alternative Versions and Optimizing the Attack}
\label{BlockWitholdAttackAlternativeVersions}

In another version of this attack
rogue miners could be total free riders:
they could also just draw money from the mining pools
and never contribute anything.
This is precisely what is called ``Sabotage'' in \cite{RosenfeldPoolRewardPaper}.
However in this case they would not obtain
gains higher in proportion than their relative computing power.
In the attack we have described they do and rogue miners
spend half of their computing power on each side.
In fact it is easy to show that this the {\bf optimal} choice.
More generally by redoing
the same analysis in the general case
we obtain immediately that:

\begin{thm}[Generalized Attack Analysis]
\label{Block WithholdingAttack}
If the proportion of rogue miners is $\alpha$ and
if a proportion of $\alpha\beta$ infiltrates the pools and discards the winning blocks,
while the other $\alpha(1-\beta)$ mine normally in a small independent pools,
then a part of rogue miners are paid more than the other miners
by the factor of $\frac{1-\alpha(1-\beta)}{1-\alpha}$ and overall
their relative gain is 
equal to:
\vskip-3pt
\vskip-3pt
$$
(1-\beta)\frac{1-\alpha(1-\beta)}{1-\alpha} + \beta - 1 =
\frac{\alpha\beta(1-\beta)}{1-\alpha}.
$$
Moreover for any given fixed $\alpha$ the gain of the rogue miners
is maximized when $\beta=1/2$.
\vskip-4pt
\end{thm}

\section{Detection And Predicted Impact Of The Block Withholding Attack}

In this section we first explain some facts about
existing mining pools, large and small.
Then we are going to apply our Theorem \ref{BitcoinMiningLaw}
to some miners, and see what it implies for them in practice.

\vskip-0pt
\vskip-0pt
\begin{table}[h!]
  \caption{Some existing pools in percentage of their hashing power
  relative to the whole bitcoin network}
    \label{MajorPoolsSize}
\vskip-0pt
\vskip-0pt
$$
\begin{array}{|c|c|c|c|}
\hline
\mbox{pool}
& percentage  \\
\hline
\mbox{BTCGuild.com}
 & 26 $\%$ \\
\hline
\mbox{GHash.io}
 & 40 $\%$ \\
\hline
\mbox{Eligius.st}
 & 10 $\%$ \\
\hline
\mbox{Bitcoin.cz}
 & 6 $\%$ \\
\hline
\mbox{Bitminter.com}
 & 5 $\%$ \\
\hline
\end{array}
$$
\vskip-1pt
\vskip-1pt
\end{table}
\vskip-0pt

In Table \ref{MajorPoolsSize} we see that bitcoin
can hardly be called a decentralized currency.
On the contrary in January 2014
just one single company based in Ukraine has
been dangerously approaching the barrier of 50 $\%$,
see \cite{GHash45Pc}).

Now we are going to show that
our attack also can contribute to centralization:
it implies a minimal size for pools
below which the attack will not be detected.

\subsection{Detection of Block Withholding}

Now armed with Theorem \ref{BitcoinMiningLaw}
we are going to show that for an individual miner
the attack is not visible.
We will look at the blocks mined by one single miner. 

We have selected one miner at random and
have observed his whole history which spans a period of 1 year.
This miner has mined $K=18$ blocks 
in one year under one single public key. 
Now following  Theorem \ref{BitcoinMiningLaw},
if one pool has produced just about 18 valid blocks in a year,
the standard deviation will be about $\sqrt{K}=\sqrt{18}\approx 4.2$,
which is about 24 $\%$ of the number $K$.
Now because due to Theorem \ref{CentralLimit} the distribution is a Gaussian,
we know that any difference between our model and the reality which is about
24 $\%$ is going to happen with very high probability close to 1, see \cite{wikistdev}.
In other words this miner
cannot detect rogue miner groups
as large as 20 $\%$ (achieving a 6 $\%$ competitive advantage)
such as described in our attack.
This is simply because any observable difference
will be way below the standard deviation
which was estimated to be about 24 $\%$.
This miner {\bf cannot possibly detect
the block withholding attack described in this paper.}

{\bf Remark 1.}
The same result holds for the majority of existing
miners with exception of very large miners.
For a miner to mine $K=18$ blocks in one year is already quite rare.
A quick estimation shows that an average miner mines less than one block per year:
BTCGuild reports a population of some 27,000 active online miners
and has a capacity to produce 
about 13,000 blocks per year.

{\bf Remark 2.}
We can observe that the bitcoin mining events which occur every 10 minutes
are {\bf not frequent enough}
for individual miners to detect rogue mining activity in practice.
If bitcoin was re-designed in such a way that the mining events
were more frequent, for example one event every second,
then in proportion the standard deviation
would be roughly $\sqrt{600}\approx 24$
times smaller for any miner or group.
This would mean that the thresholds at which rogue
activity could be detected by miners
would be roughly also about 24 times smaller.
However for the time being individual bitcoin miners just
cannot detect these important and realistic attacks, not even in theory.

{\bf Remark 3.}
If all bitcoin miners in existence published their detailed mining stats
and their exact numbers of trials,
rogue miner strategies would be detected more easily and with better precision.
Large mining pools will be large enough to detect the attack.
However the miners must trust the pool to detect the attack and
to report the statistical data accurately.
Mining pools could have an interest in concealing the existence of the attack.
Maybe because some strategies against Pool Hopping
attacks \cite{RosenfeldPoolRewardPaper,RosenfeldPoolRewardMethods}
involve concealing information about the mining process. 
More importantly, in order to avoid users from switching to another pool
which could be in some cases more harmful than the actual attack
(in both cases the pool income would decrease).

%
%
%
%

\subsection{Have Rogue Strategies Been Applied?}
\label{HaveRogueBeenDone}

We have looked at public stats presented by the second largest pool in existence, BTCGuild.com.
Have block withholding already been done?
In order to see this we have multiplied the currency difficulty level by
the overall Pay Per Share rate displayed by BtcGuild.com on 05/01/2014. We obtain.

\vskip-4pt
\vskip-4pt
$$
1418481395*0.0000000163026460= 23.125
$$
\vskip-4pt

The result should be at least 25 bitcoins
plus transaction fees, minus 5+3 $\%$ pool fees
which are kept by this pool according to \cite{BitcoinExistingPools}.
This should be at least 23 BTC plus some 0.1-0.2 BTC due to typical transaction fees.
This is more or less what we obtain.

We have also looked at some individual miners.
Only for very large miners we can expect to see anything, cf. Thm. \ref{BitcoinMiningLaw}.
The biggest individual miner 269032 had the hashing power of 174 TH/s on 01/01/2014.
A quick calculation shows that he should produce 2.5 blocks/day on average.
In five days this will be some 12.5 blocks on average with a standard deviation of about 3.5
cf. Thm. \ref{BitcoinMiningLaw}.
We have observed in the public stats that this user has produced 16 blocks in the first 5 days of 2014
and 10 blocks in the following 5 days. These figures fit perfectly within the standard deviation interval.

Overall we conclude that there is no evidence that rogue miner strategies have ever been applied.

\medskip
\medskip
{\bf LATEST NEWS 13 JUNE 2014}:
It seems that a block withholding attack
was executed in practice
against Eligius mining pool,
see Section \ref{RecentWithholdAttackEligius300BTC}.

{\bf Earlier attacks against 50 BTC}:
There were also earlier reports of {\em suspected} block withholding
by an 'evilpool' acting as a proxy to
connect to 50BTC mining pool in late 2013,
cf. \cite{PoolServerHacksRedirectionWithholding2013},
in which the 'evilpool' would send to 50BTC pool
"all shares except of winning ones",
see \cite{PoolServerHacksRedirectionWithholding2013}.

\newpage
~
\newpage

\vskip-4pt
\vskip-4pt
\section{Conclusion}
\label{sec:conclusion}
\vskip-1pt

Bitcoin digital currency is a sort of distributed electronic notary system.
It is used to 
make payments with units called bitcoins.
It has a public ledger of all transactions which is used to record all events
in which money circulates between different participants
identified by their cryptographic keys.
Bitcoin miners are the key people
which support this digital currency by participating in a
sort of lottery which attributes important monetary rewards to them.
Their activity has substantial computational cost
which is the main thing which makes it difficult for the
attacker to modify the common history of monetary transactions.
In this paper we look at the question of subversive strategies for miners
in which they deviate from the expected ``honest'' 
behavior. We survey recent results and propose new attacks.

Bitcoin cultivates this ``impossible'' cryptographers' dream
of building a payment system without any trusted parties.
In theory bitcoin is a network which is expected to police itself.
For example miners not following the protocol risk that their blocks
will be later rejected be the majority of other network participants
and they will not receive the reward which usually motivates them to support this digital currency.
In practice however as we show in this paper,
not every rogue behavior is detectable.
If a certain form of unwanted behavior is not visible,
it can hardly be policed or prevented.


In this paper we looked at a number of different 
methods by which some participants in the bitcoin digital currency
can hope to increase their gains at the expense of others.
Several such attacks have been proposed in the past
\cite{CartelAttack,RosenfeldPoolRewardPaper,RosenfeldPoolRewardMethods,BlockDiscardingNetworkSuperiority}.
Many more such attacks can be proposed,
for example any non-trivial cryptographic
optimisation in the bitcoin mining process
could be kept confidential by a group of miners,
cf. Section \ref{CryptoOptimizationAttack}.
A very recent method is the so called selfish miner strategy
from \cite{BitcoinBrokenGrowingSelfishPoolStrategy} which
was also independently proposed in \cite{BlockDiscardingNetworkSuperiority}.
In this paper we question the assumptions of \cite{BitcoinBrokenGrowingSelfishPoolStrategy}
and we have come to a temporary conclusion that
this strategy is unlikely to make any difference in practice
because it depends on events in the bitcoin network are visible
and which have been historically excessively rare.
As such they are unlikely to substantially
increase in the future.
For example because one method to reform the behavior of miners
was precisely already proposed in \cite{BitcoinBrokenGrowingSelfishPoolStrategy}
and a method to discourage forks have been proposed in \cite{BlockDiscardingNetworkSuperiority}.

Another major idea which was already proposed in the past are subversive
strategies which involve block withholding \cite{RosenfeldPoolRewardPaper,BitcoinBrokenGrowingSelfishPoolStrategy}.
In this paper
we have described and analyzed
a specific practical block withholding attack
which to the best of our knowledge is a new and original block withholding attack.
It generalizes the ``Sabotage'' attack of \cite{RosenfeldPoolRewardPaper}
and achieves superior gains for the subversive miners.
The two main results of \cite{BitcoinBrokenGrowingSelfishPoolStrategy} are as follows.
(1) to design an attack in which the rogue miners will obtain gains which
in proportion to their contributed computing power
will be higher than for the honest majority,
and (2) to show that this creates substantial incentives for
greater centralization in the bitcoin network
which is somewhat contrary to the whole idea of bitcoin.
In this paper we also achieve (1) and (2) by a different and independent method.
%
Moreover we show 
that our strategy can tolerate a substantial percentage
of miners engaged in it,
for example as big as 20 $\%$,
this without being detected by small pools or individual miners.
%
In contrast the so called selfish miner strategy
of \cite{BitcoinBrokenGrowingSelfishPoolStrategy}
depends on events in bitcoin history (forks)
which are always visible and can hardly be ignored.
In comparison we claim that the block withholding attacks are impossible to avoid
otherwise than by {\bf trusting} the miners who participate in pools
which is disturbing knowing that bitcoin was designed to precisely avoid
trusting other network participants.
The only defense against this attack we can see is that pools should involve ONLY people
which they personally know and trust,
and the pool manager should simply dissolve and close a pool as soon
as he notices that it is earning less than expected from its computational effort.
This leads to a postulate that pools should not be too large.
On the other hand, pools cannot be too small,
otherwise miners will cheat without being detected,
due to Thm. \ref{BitcoinMiningLaw}.


In practice
it is well known that there are strong incentives for miners to
``flock to the biggest pools'' \cite{RosenfeldPoolRewardMethods}.
Contrary to the popular wisdom,
bitcoin is not really decentralized.
Bitcoin mining is very highly centralized
and recently one pool was approaching
the half of the computing power in the bitcoin network, cf. Table \ref{MajorPoolsSize}.
This is a big problem and it contradicts the original idea
of bitcoin as a decentralized currency 
\cite{SatoshiPaper}.
In this paper we provide additional reasons for miners to avoid small pools
and provide another plausible explanation
why such
a high concentration of hashing power is natural and very hard to avoid.
We postulate that the subversive mining strategies
might in the future play a positive role in the
bitcoin network:
they might lead to miners avoiding large pools
because it will be shown that these pools contain subversive miners
earning more than their fair share.
At this moment in history there is no evidence of any rogue miner behavior.
Yet people mine in excessively large ``monopoly'' pools even
when they have 
very high fees as big as 8 $\%$, see Section \ref{HaveRogueBeenDone}.
This shows that various recent results about optimal strategies in bitcoin network remain very academic.
Many people will just {\bf not} change their behavior in order to achieve a gain of a few percent.



\medskip

\subsection{Recent Developments - June 2014}
\label{RecentWithholdAttackEligius300BTC}

Early reports of {\em suspected} block withholding attacks
go back to late 2013,
cf. \cite{PoolServerHacksRedirectionWithholding2013}
and Section \ref{HaveRogueBeenDone}.

On 13 June 2014 it was reported that
{\bf a large-scale block-withholding attack }
as described in this paper (or a variant)
{\bf was executed against the mining pool Eligius},
see
\url{https://bitcointalk.org/?topic=441465.msg7282674}.
{\bf Losses are very substantial and were estimated
to be about 300 BTC} at the expense of honest miners,
The pool has been able to detect the attack
and was able to block 200 BTC worth of the attackers payouts'
which from the point of view of Eligius
were as good as stealing 200 bitcoins earned by the honest miners.
The attackers have in retaliation
"threatened putting a 200 BTC bounty on hacking Eligius".

In the same blog post we read:
{\bf
"the attacker does not gain any direct benefit by performing the attack". }
This reflects the general lack of understanding
of such attacks in the bitcoin community.
Also the Cornell researchers have claimed it is NOT profitable: 
"Note that the attacker doesn't gain anything from this behavior, either;
{\bf it's purely destructive"},
see \cite{CornellTimeForHardForkAvoid51GHashWithholdingEtc}.

{\bf
In this paper we show that this attack CAN be profitable }
and we believe that the attack executed
against Eligius was run in such a way as to be profitable,
possibly exactly as described in this paper.
This is likely to be true given the fact
that the variant we describe in this paper
was optimized to maximize the profitability.

{\bf Recent Developments:} 
A new paper takes block withholding attacks 
to the new level \cite{EyalMinersDilemmaNov2014} (26 November 2014). 
This recent work is build upon one of the key observations in the present paper, 
which is that miners can withhold block selectively in some pools, 
and mine normally in other pools. 
This is 
exactly what can make the block withholding attacks eventually profitable 
(cf. our present paper) and not just "purely destructive"
cf. \cite{CornellTimeForHardForkAvoid51GHashWithholdingEtc}. 
However these attacks are really profitable mainly 
if one miner executes them, and other miners don't. 
The new paper considers further more complex scenarios 
where several miners are trying to cheat simultaneously, 
which decreases the incentives for the attack 
and potentially might convince the miners to be honest. 

The new paper also claims that this "would push miners to join 
private pools which can verify that their registered miners do 
not withhold blocks". This is not very likely. 
No pools can detect the attack if it is done correctly. 
We recall that block withholding attacks are very hard to detect
cf. Section \ref{BitcoinMiningLawSection} 
unless the attacker is not very careful and mines large quantities of bitcoins under the same address,
cf. Section \ref{HaveRogueBeenDone}. 
The attacker can easily execute a block withholding attack against any 
set of pools in such a way that the pools will not be able to incriminate 
a single address which belongs to the attacker, for this it is sufficient to 
fragment the attack and mine under many different identities. 

\medskip
{\bf Acknowledgments:}
We thank
Alex Biryukov for his extremely helpful
suggestions 
and comments.


\newpage

\newpage
\centerline{APPENDIX}

\section{The Good and the Ugly:
Subversive Strategies For ASIC Manufacturers}

In bitcoin network there exists a threat which is far bigger than
some malicious miners and centrally managed mining pools.
In this paper we have seen that bitcoin mining
is surprisingly centralized instead of being decentralized,
cf. Table \ref{MajorPoolsSize}.
In this section we look at the problems of trust and risk of fraud
which result from the highly centralized supply of mining devices.

In 2012-2013 
a new powerful group of people have emerged in the bitcoin currency:
ASIC miner chip manufacturers.
A dozen of companies which offer fast
specialized ICs for bitcoin mining.
These devices
achieve power consumption which is up to 10,000 times smaller
than what can be achieved with ordinary computers 
cf. 
\cite{MiningUnreasonable}.
At this moment bitcoin has totally lost
whatever remained from a peer-to-peer utopia without trusted parties.
A dozen of companies control 100 $\%$ of bitcoin mining devices market.
These companies are the lifeblood of the bitcoin currency.
They must be trusted and trustworthy for bitcoin to function.
They take pre-orders and manipulate tens of millions of dollars of other people's money.
Unhappily these companies are subject to important moral hazards as we will see below.

In the traditional financial sector,
companies which might be tempted to misbehave 
are strictly controlled and regulated by the governments.
In bitcoin we have a unique situation where these companies function
as ordinary businesses which is very unusual.
In this respect bitcoin is not like other currencies.
We live in a society of {\bf excessive asymmetry of information} between
the public and the private sector. It is very easy to get information about actions of
governments and politicians, and it is very common to criticize their actions. 
In the recent years it also became possible to criticize the government secret services such as the NSA, 
and since the financial crisis of 2008 it became very common to publicly ostracize banks and financial institutions.
However it is very rare and quite difficult to criticize private business.
Firstly this is because it is extremely difficult to get accurate data on their actions.
Secondly almost nobody ever does it fearing legal proceedings.
However this should not stop researchers from investigating these questions.
Security researchers need to study all sort of threats to the secure systems
and many issues should be discussed publicly if we want the security to improve.
A market economy cannot function properly if customers cannot know
what kind of company they are ordering things from.

Below we give a short summary of dirty tricks which ASIC manufacturer companies
could be tempted to execute on their customers.
It is not our role to accuse any of these companies.
Moderate delays in delivery are not surprising for a new ASIC,
and we believe that most of these companies are honest,
hard working and ethical businesses.
Cf.
\footnote{
\label{foot2}
We have however observed that the Internet is full of reports of angry customers
of the US company ButterFly Labs, which has excessively bad reputation
and has fallen very badly below customer expectations
with delivery delays as long as 1 year.
} 
$^{\ref{foot3}~\ref{foot4}}$.
%
The competition between these companies have recently become quite fierce
and hopefully this creates strong incentives to behave well.
However.
When miners invest money in order to support
a digital currency 
they must be aware of the risks.
Here are the principal threats
and attacks we have identified:


\begin{enumerate}
\item {\bf Non-existent Production Attack}:
The miners advertise and sell mining devices which do not exist.
Payment is done in bitcoins only (bad sign).
Cf. 
\footnote{
\label{foot3}
Some companies are clearly fake
and do not exist, for example {\bf \url{www.hashblaster.com}}
and
{\bf \url{www.xtrememiners.net}}.
They are just criminals who take the money and provide nothing in return.
}
\footnote{
\label{foot4}
There is a web site which reports all sort of bitcoin
fraudulent activity:
{\bf \url{http://bitcoinscammers.com}}
such as fake miner stores, for example
{\bf \url{minerstore.net}}~and~{\bf \url{ctsminer.com}} etc.
are all scams.
}.
\item {\bf Fake Retailer Attack}:
Devices are sold by a third-party retailer
which does not deliver miners to customers,
just collects the money.
Cf. $^{\ref{foot4}}$.
\item {\bf Secret Production Attack}:
ASIC manufacturers conceal how many devices have been ordered
in order to make miners believe that mining will be profitable,
which allows to sell even more such devices.
\item {\bf Device Hoarding Attack}:
The ASIC manufacturer delays shipping the device for a very long time like 1 year without a reason.
He uses the device to mine bitcoins and makes a lot of money at the expense of a naive customer.
\item {\bf Power Consumption Attack}:
The ASIC manufacturer advertises a power consumption of 1 W per Gigahash/s and is so confident about achieving this target
that he offers to pay a very large lump sum of money to charity if they miss this target by more than 10 $\%$.
Many people order these devices and the ASIC company
is able to capture a lot of cash from miners worldwide.
Later customers will receive devices which consume 3.2 Watts per Gigahash/s.
All is blamed on technical problems and lack of experience of the manufacturer.
In fact it could have been a deliberate strategy to get more orders.
\item {\bf Device Substitution Attack}:
The ASIC manufacturer remains very vague on specifications of their devices,
and never specifies the power consumption in the commercial material.
Then he ships inferior products to most customers.
\item {\bf Device Backdoor Attack}:
The ASIC manufacturer ships devices with nominal power of 3.2 W however they are able to manufacture an IC consuming 1 W.
The remaining 2.2 W are spent on a hidden functionality: mining for the rogue manufacturer.
Only when a block is found the backdoor functionality must communicate with the manufacturer.
This s a very rare event and the fraud remains undetected.
The ASIC company is able to mine bitcoins in vast quantities with {\bf both} hardware and electricity
being paid by naive miners, and this without being detected. In addition the ASIC manufacturer can implement
anti-reverse engineering countermeasures which are well known in the silicon industry,
to avoid the fraud from being detected.
\item {\bf Mafia Lottery Attack}:
The ASIC manufacturer announces that customers which are going to receive devices first are going to be selected by a lottery.
This can be seen as either a really brilliant method to boost sales
or a dubious and rather terribly immoral policy,
this depending on the point of view.
It is 1) a slap in the face of customers who paid earlier
and 2) an incitation for even more customers to order devices,
because they have some chance to obtain them at the expense of customers who paid earlier.
Now here is what is (potentially) going on behind the scenes.
We describe just one of a few possible subversive scenarios.
In addition to the official market far selling ASIC devices,
there is another parallel market where people can buy something else for example shares
of some company or hosted mining services.
However in fact the lottery is rigged,
and the second market functions as a method to collect kickbacks for the first market,
while the first market allows to artificially inflate assets on the second market.
The company makes a lot of money on both markets
and at the same time they claim that they are doing nothing wrong.
In fact the fact alone of using a lottery should be a warning sign.
\end{enumerate}

\end{document}